\documentclass[11pt]{article}
\usepackage{moriond,epsfig}
\usepackage{amsmath,amssymb}
\usepackage[all]{xy}
\usepackage{slashed}
\usepackage{psfrag}

\def\Journal#1#2#3#4{{#1} {\bf #2}, #3 (#4)}

\def\be{\begin{equation}}
\def\ee{\end{equation}}
\def\bea{\begin{eqnarray}}
\def\eea{\end{eqnarray}}

\newcommand{\rxy}[1]{{\begin{xy}0;<2mm,0mm>:<0mm,2mm>::0;0,#1\end{xy}}}

\newcommand{\minibox}[2]{{\fbox{\begin{minipage}{#1}\begin{center}#2
\end{center}\end{minipage}}}}

\begin{document}
\vspace*{4cm}
\title{BEYOND THE STANDARD MODEL: A NONCOMMUTATIVE APPROACH}

\author{ C.A. STEPHAN }

\address{Institut f\"ur Mathematik, Universit\"at Potsdam\\
14469 Potsdam, Germany}

\maketitle\abstracts{
During the last two decades Alain Connes developed Noncommutative Geometry 
(NCG)\cite{book},  which allows to unify two of the basic theories of modern physics: General Relativity (GR) and the Standard Model (SM) of Particle Physics as classical field theories\cite{cc}. In the noncommutative framework
the Higgs boson, which had previously to be put in by hand,  and many of the ad hoc features of the standard model appear in a  natural way. 
The aim of this presentation is to motivate this unification from
basic physical principles and to give a flavour of its derivation. One basic prediction
of the noncommutative approach to the SM is that the mass of the
Higgs Boson should be of the order of 170 GeV if one assumes the Big Desert\cite{cc}. This mass range is with reasonable probability excluded by the
Tevatron\cite{TEV} and therefore it is interesting to investigate 
models beyond the SM that are compatible with NCG. 
Going beyond the SM is highly non-trivial within the NCG approach
but  possible extensions have been found\cite{Be,Scal} and provide for 
phenomenologically interesting models. We will present in this article a
short introduction into the NCG framework and describe 
one of these extensions of the SM. This model contains 
new scalar bosons 
(and fermions) which constitute a second Higgs-like sector mixing with the
ordinary Higgs sector and thus considerably
modifying the mass eigenvalues.}

\section{Noncommutative Geometry}
The aim of NCG: To unify GR and the SM (or suitable extensions of the SM) 
on the same geometrical footing. This means to describe gravity
and the electro-weak and strong forces as gravitational forces of a unified space-time. 
\\ \\
First observation, the structure of GR:
gravity emerges as a pseudo-force associated to the space-time (=  a manifold $M$) symmetries, i.e.
the diffeomorphisms on $M$.
If one tries to put the SM into the same scheme, one cannot find a manifold 
within classical differential geometry 
which is equivalent to space time.

Second observation: one can find an equivalent description of space-time
when trading the differential geometric description of the Euclidean 
space-time manifold $M$ with metric
$g$ for the
algebraic description of a spectral triple $(C^\infty(M),
\slashed{\partial},\mathcal{H})$. A spectral triple consists of the following
entities:
\begin{itemize}
\item[$\bullet$] an algebra $\mathcal{A}$ $(=C^\infty(M))$, the equivalent of the topological space $M$
\item[$\bullet$] a Dirac operator $\mathcal{D}$ $(=\slashed{\partial})$, the equivalent of the metric $g$
on $M$
\item[$\bullet$] a Hilbert space $\mathcal{H}$, on which the algebra is faithfully represented and on which the Dirac operator acts. It contains the fermions, i.e. in
the space-time case, it is the Hilbert space of Dirac 4-spinors.
\item[$\bullet$] a set of axioms \cite{book}, to ensure a consistent description of
the geometry
\end{itemize}
This allows to describe GR in terms of spectral triples. Space-time is replaced by the
algebra of $C^\infty$-functions over the manifold and the Dirac operator plays a double role: 
it is the algebraic equivalent of the metric and it gives the dynamics of the Fermions.

The Einstein-Hilbert action is replaced by the spectral action \cite{cc}, which is the simplest invariant action, given by the number of eigenvalues of the Dirac operator up to a cut-off. 
\begin{center}
\begin{tabular}{c}
\\
\rxy{
,(-20,15)*{
\minibox{3cm}{Space-Time (S-T)\\ 4-dim Manifold $M$}
}
,(20,15)*{
\minibox{3cm}{S-T Symmetries \\ Diffeomorphisms of $M$}
}
,(-9,15);(9,15)**\dir{-}?(0)*\dir2{<}
,(-20,9);(-20,4)**\dir3{-}
,(20,9);(20,4)**\dir3{-}
,(-14,7)*{\mbox{equiv.}}
,(26,7)*{\mbox{equiv.}}
,(0,18)*{\mbox{act on}}
,(-20,0)*{
\minibox{3cm}{Spectral Triple\\ $(C^\infty(M),\slashed{\partial},
\mathcal{H})$}
}
,(20,0)*{
\minibox{3cm}{ Automorphisms \\ of $C^\infty(M)$}
}
,(0,-17)*{
\minibox{4cm}{Gravity \\ Spectral Action = \\ E-H Act.+Cosm.Const.}
}
,(-20,-5);(-1,-11)**\dir{-}?(0)*\dir2{<}
,(20,-5);(1,-11)**\dir{-}?(1)*\dir2{>}
,(-9,0);(9,0)**\dir{-}?(0)*\dir2{<}
,(0,3)*{\mbox{act on}}
,(24,-8)*{\mbox{leave invariant}}
,(-20,-8)*{\mbox{Dynamics}}
}
\\ \\
\end{tabular}
\end{center}
Connes' key observation is that the geometrical notions of the spectral triple remain
valid, even if the algebra is {\it noncommutative}. A simple way of achieving
noncommutativity is done by multiplying the function
algebra $C^\infty (M)$  with  a
sum of matrix algebras
\begin{equation}
\mathcal{A}_f = M_1(\mathbb{K}) \oplus M_2(\mathbb{K}) \oplus \dots
\label{intalg}
\end{equation}
with $\mathbb{K}=\mathbb{R}$, $\mathbb{C}$ or $\mathbb{H}$. These algebras,
or {\it internal spaces}
have exactly the $O(N)$-, $U(N)$- or $SU(N)$-type Lie groups as their symmetries
that one would like to obtain as gauge groups in particle physics. The particular
choice
\begin{equation}
\mathcal{A}_f = \mathbb{C} \oplus M_2(\mathbb{C}) \oplus M_3(\mathbb{C})
\oplus \mathbb{C}
\label{SM}
\end{equation}
allows to construct the SM. The combination of space-time and internal space
can be considered as a {\it discrete Kaluza-Klein space} where the 
product-manifold of space-time with extra dimensions is replaced
by a product of space-time with discrete spaces represented by matrices,
see Fig. 1. These discrete spaces allow just for a finite number of 
degrees of freedom and  avoids the infamous Kaluza-Klein 
towers. 
\begin{figure}
\begin{center}
\includegraphics[width=6cm]{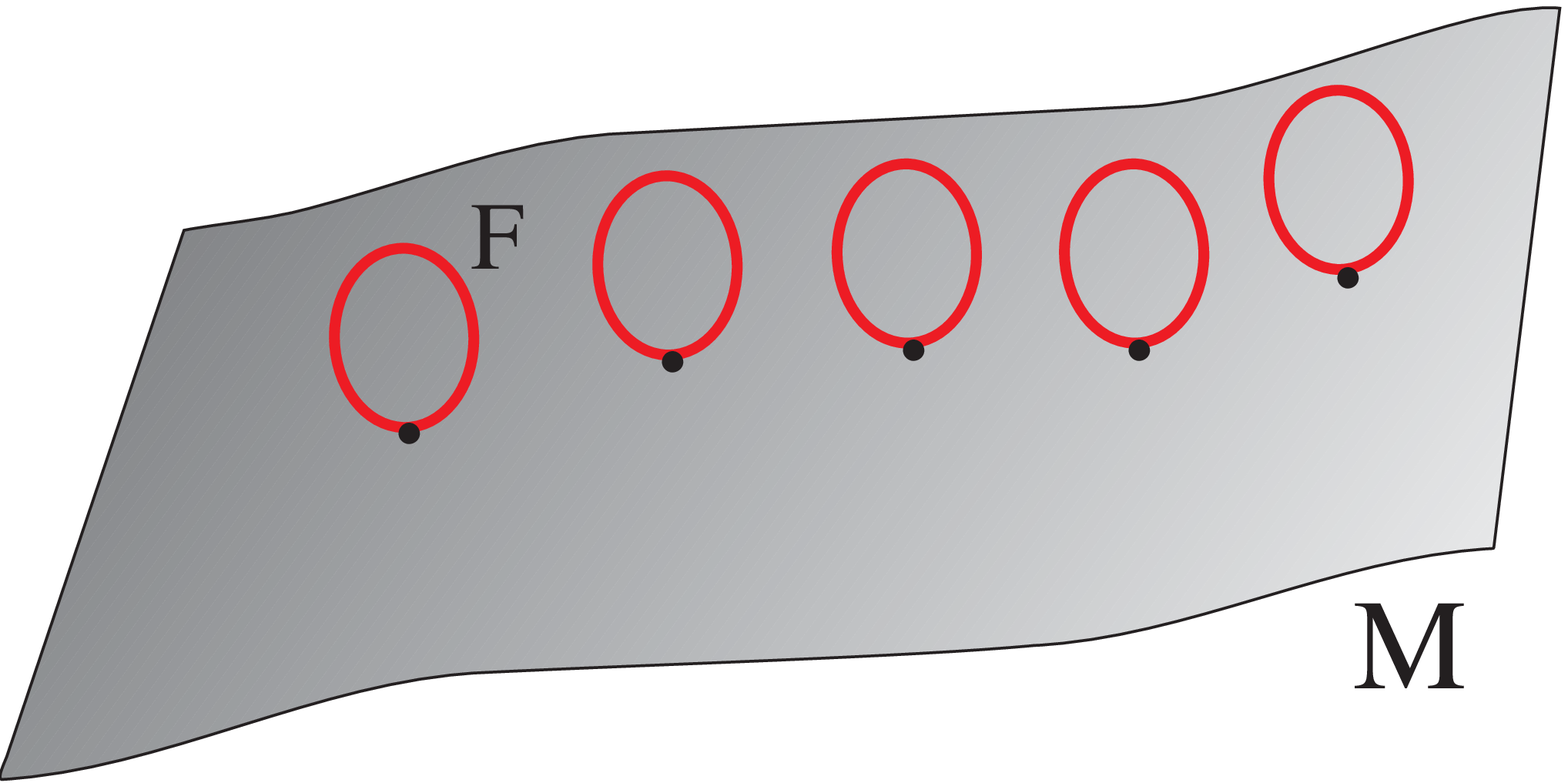} 
\makebox(80,80){ 
\includegraphics[width=3cm]{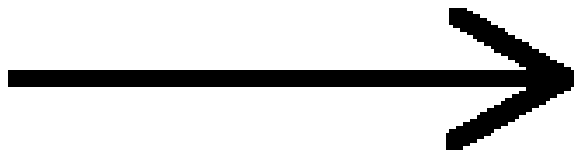} }
\psfrag{A}{$C^\infty(M)$}
\psfrag{B}{$\mathcal{A}_f$}
\includegraphics[width=6cm]{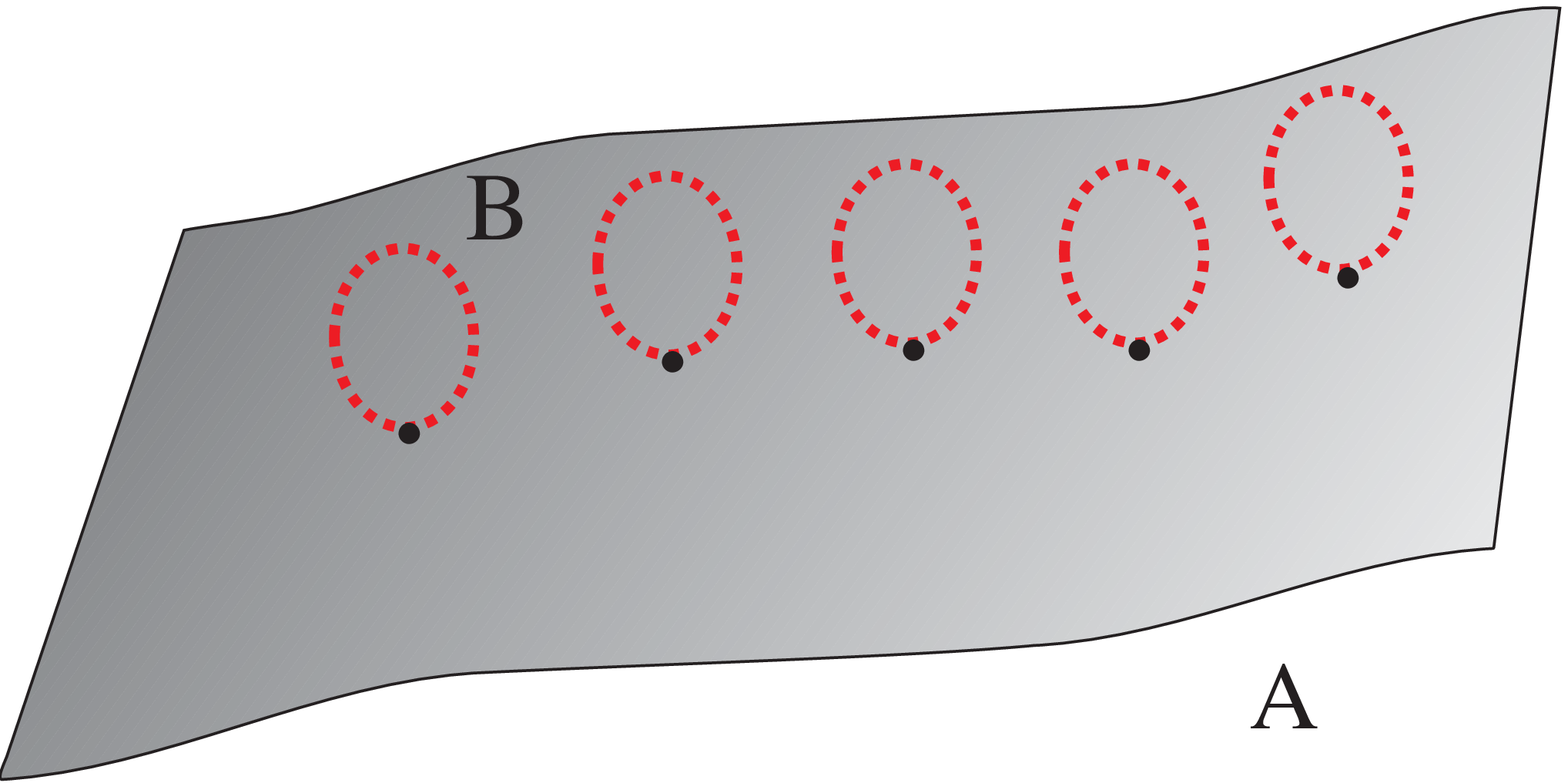} 
\end{center}
\caption{A noncommutative geometry can be pictured by 
imagining a Kaluza-Klein space where the compact extra dimensions
$F$ are replaced by discrete matrix spaces $\mathcal{A}_f$ and
space-time $M$ by the algebra of smooth functions $C^\infty(M)$ over $M$.}
\end{figure}
 \\ \\
These product spaces 
the spectral triple approach, combined with the spectral action \cite{cc},
unify GR and the SM as classical field theories:
\begin{center}
\begin{tabular}{c}
\\
\rxy{
,(-20,0)*{
\minibox{4cm}{Almost-Commutative \\ Spectral Triple \\ $\mathcal{A}= C^\infty(M) \otimes
\mathcal{A}_f$}
}
,(20,0)*{
\minibox{4cm}{Int.+Ext. Symmetries \\ = Diff$(M) \ltimes$ \\ $U(1) \times SU(2) \times SU(3)$}
}
,(0,-20)*{
\minibox{5cm}{Spectral Action = \\ E-H Act.+Cosm.Const. \\ + Stand. Model Action }
}
,(-15,-6);(-1,-15)**\dir{-}?(0)*\dir2{<}
,(15,-6);(1,-15)**\dir{-}?(1)*\dir2{>}
,(-9,0);(9,0)**\dir{-}?(0)*\dir2{<}
,(0,3)*{\mbox{act on}}
,(18,-10)*{\mbox{leave invariant}}
,(-14,-10)*{\mbox{Dynamics}}
}
\\ \\
\end{tabular}
\end{center}
The  SM in the NCG setting automatically produces:
\begin{itemize}
\item[$\bullet$] The combined GR and SM action
\item[$\bullet$] A cosmological constant
\item[$\bullet$] The Higgs boson with the correct quartic Higgs potential
\end{itemize}
The Dirac operator turns out to be one of the central objects and plays a
multiple role:

\begin{center}
\begin{tabular}{c}
\rxy{
,(0,1)*{\minibox{2.5cm}{Dirac Operator}}
,(-25,-15)*{\minibox{2.6cm}{Higgs \& Gauge \\ Bosons}}
,(25,-15)*{\minibox{2.6cm}{Metric of $M$,\\ Internal Metric}}
,(0,-15)*{\minibox{3.5cm}{Particle Dynamics, \\ Ferm. Mass Matrix}}
,(-1,-3);(-25,-9)**\dir{-}?(1)*\dir2{>}
,(1,-3);(25,-9)**\dir{-}?(1)*\dir2{>}
,(0,-3);(0,-9)**\dir{-}?(1)*\dir2{>}
}
\\ \\
\end{tabular}
\end{center}

\section{Physical Consequences of the NCG formalism}

Up to now one could conclude that NCG consists merely in a fancy, 
mathematically involved reformulation of the SM.
But the choices of possible Yang-Mills-Higgs models that fit into
the NCG framework is limited. Indeed the geometrical setup leads
already to a set of restrictions on the possible particle models:
\begin{itemize}
\item mathematical axioms  $=>$ restrictions on particle content
\item symmetries of finite space $=>$ determines gauge group
\item representation of matrix algebra  $=>$ representation of gauge group

(only fundamental and adjoint representations)
\item Dirac operator $=>$ allowed mass terms / Higgs fields
\end{itemize}
Further constraints come from
the Spectral Action which results in an effective action valid at a
cut-off energy $\Lambda$  \cite{cc}. This effective 
action comes with a set of constraints on the standard model
couplings, also valid at $\Lambda$, which reduce the 
number of free parameters in a significant way.

The Spectral Action contains two parts. A fermionic part $( \Psi, \mathcal{D} \Psi )$
which is obtained by inserting the Dirac operator into the scalar product
of the Hilbert space and a bosonic part $S_\mathcal{D} ( \Lambda)$
which is just the number of eigenvalues of the Dirac operator up to
the cut-off:
\begin{itemize}
\item $( \Psi, \mathcal{D} \Psi )$ = fermionic action  includes Yukawa couplings 
 \& fermion--gauge boson interactions
\item $S_\mathcal{D} ( \Lambda)$ = the bosonic action given by the
 number of eigenvalues of $\mathcal{D}$ up to cut-off $\Lambda$

\hskip1.2cm = the Einstein-Hilbert action + a Cosmological  Constant 

\hskip1.5cm + the full bosonic SM action + constraints at $\Lambda $

\end{itemize}
The bosonic action $S_\mathcal{D} ( \Lambda)$ can be calculated explicitly 
using the well known heat kernel expansion \cite{cc}. Note that $S_\mathcal{D} ( \Lambda)$ is manifestly gauge invariant and also invariant under
the diffeomorphisms of the underlying space-time manifold.

For the SM the internal space is taken to be the matrix algebra
 $\mathcal{A}_f=  \mathbb{C} \oplus M_2(\mathbb{C} ) \oplus M_3(\mathbb{C} )
\oplus \mathbb{C}$. The symmetry  group of this discrete space
is given by the group of (non-abelian) unitaries of $\mathcal{A}_f$: 
$U(2)  \times U(3)$. This leads, when properly lifted to the Hilbert space,
to the SM gauge group  $U(1)_Y \times SU(2)_w  \times SU(3)_c$.
It is a remarkable fact that the SM fits so well into the NCG framework!

Calculating from the geometrical data the Spectral Action leads to the
following constraints on the SM parameters at the cut-off $\Lambda$
\begin{equation}
 g_2^2=\, g_3^2=\,\frac{Y_2^2}{H} \,\frac{\lambda}{24}\,= \,\frac{1}{4}\,Y_2 
\end{equation}
Where  $g_2$ and $g_3$ are the $SU(2)_w$ and $SU(3)_c$  gauge couplings,
$\lambda$ is the quartic Higgs coupling,
$Y_2$ is the sum of all Yukawa couplings squared
and
$H$ is the sum of all Yukawa couplings to the fourth power.

Assuming the Big Desert and the stability of the theory under the
flow of the renormalisation group equations we can 
deduce that $g_2^2(\Lambda) =g_3^2 (\Lambda) $ at $\Lambda = 1.1 \times 10^{17}$ GeV (where we used the experimental values  $g_{2}(m_Z) =0.6514$, $\; g_{3}(m_Z)=1.221$). Having thus fixed the cut-off scale $\Lambda$ we can
use the remaining constraints to determine the low-energy value
of the quartic Higgs coupling and the top quark Yukawa coupling (assuming
that it dominates all Yukawa couplings). 
This leads to the following conclusions:
\begin{itemize}
\item $m_{\rm Higgs} = 168.3 \pm 2.5$ GeV
\item $m_{\rm top} < 190$ GeV
\item no $4^{th}$ SM generation
\end{itemize}
Note that the prediction for the Higgs mass is already problematic
in the light of recent data from the Tevatron \cite{TEV}. Thus it seems
plausible to consider also models beyond the SM. We will present 
one recently discovered model which may have an interesting 
phenomenology. 

\section{Beyond the Standard Model}

The general strategy to find models beyond the SM within
the NCG framework can be summarised as follows:
\begin{itemize}
\item find a finite geometry that has the SM as a sub-model (tricky)

$=>$ particle content, gauge group \& representation
\item (make sure everything is anomaly free)
\item compute the spectral action $=>$ constraints on the parameters at $\Lambda$
\item determine the cut-off scale $\Lambda$ with suitable sub-set
of the constraints
\item use renorm. group equations to obtain the low energy values
of interesting parameters 
\item check with experiment!
\end{itemize}
This general strategy led, among other models \cite{Be}, to the following extension
of the SM \cite{Scal}: The model contains in addition to the SM particles 
a new scalar field and a new fermion singlet in each generation.
These particles are neutral with respect to the SM gauge group
but charged under a new $U(1)_X$ gauge group under which
in turn the SM particles are neutral. 

The discrete space is a slight extension of the SM space, $\mathcal{A}_f =\mathbb{C} \oplus M_2(\mathbb{C}) \oplus M_3(\mathbb{C}) \oplus 
\mathbb{C} \oplus \mathbb{C} \oplus \mathbb{C}$, which results 
in a $U(1)$ extension of the SM gauge group: 
$G= U(1)_Y \times SU(2)_w \times SU(3)_c \times U(1)_X$.
The new fermions carry a  $U(1)_X$ charge which is vector-like  (we call them $X$-particles),
with masses  $M_X \sim \Lambda$. Note that very similar models have been considered
before \cite{Bij}.
The new scalar  $\varphi$ also transforms under  $U(1)_X$ and couples to the SM Higgs field $H$:
\begin{equation}
V_{\rm scalar} =   -\mu_1^2 |H|^2  +
\frac{\lambda_1}{6} \, |H|^4 -\mu_2^2 | \varphi|^2 +\frac{\lambda_2}{6} \, 
| \varphi|^4 +\frac{\lambda_3}{3} \, |H|^2 | \varphi|^2
\end{equation}
This Lagrangian leads to the following symmetry breaking pattern for 
the model:
\begin{equation}
U(1)_Y \times SU(2)_w \times SU(3)_c \times U(1)_X \rightarrow 
U(1)_{el.} \times SU(3)_c.
\end{equation}
For the $X$-particles and for the $U(1)_X$ gauge bosons one finds
the standard Lagrangian:
\begin{equation}
\mathcal{L}_{\rm ferm+gauge} 
=   \bar X_L M_X X_R + g_{\nu,X} \bar \nu_R
 \varphi X_L + \ h.c. \ + \frac{1}{g_4^2} F^{\mu \nu}_X F_{X, \mu \nu} 
\end{equation}
Calculating the Spectral Action leads to a set of constraints
very similar to the SM constraints,
\begin{equation}
g_2^2 = g_3^2 =  \frac{\lambda_1}{24} \frac{(3g_t^2 + g_\nu^2)^2}{3g_t^4 + g_\nu^4} =  \frac{\lambda_2}{24} =  \frac{\lambda_3}{24} \frac{3g_t^2 + g_\nu^2}{g_\nu^2} \frac{1}{4}\, (3g_t^2 + g_\nu^2),
\end{equation}
where for the sake of simplicity we have only kept the top quark Yukawa coupling
$g_t$ and the $\tau$-neutrino Yukawa coupling $g_\nu$. The latter can
be of order one and leads via the seesaw mechanism to a small neutrino mass.
The relevant free parameters in this model are the vacuum expectation 
value $|\langle \varphi \rangle|$ of the new scalar field $\varphi$ and
the $U(1)_X$ gauge coupling  $g_4$ at given energy (for example at the $Z$-mass).

One notes that the potential  $V_{ \rm scalar}$ of the scalar sector
has a non-trivial minimum. But the vacuum expectation
values of the new scalar $\varphi$ and the Higgs $H$ do no longer constitute
the mass eigenvalues. The physical scalar fields mix, thus leading to 
a dependence of the mass eigenvalues on the vacuum expectation value of
the new scalar,  $|\langle \varphi \rangle|$. The vacuum expectation value
of the SM Higgs field is still dictated by the $W$-boson mass.

Two examples of the dependence of the mass eigenvalues on  
$|\langle \varphi \rangle|$ are depicted in Fig. 2, for the exemplary
values of the $U(1)_X$ 
 gauge coupling,  $g_4(m_Z)=0.3$ (left figure) and 
$g_4(m_Z)=0.7$ (right figure). For a detailed
analysis of this model we refer to \cite{Scal}.

\begin{figure}
\begin{center}
\psfrag{AA}[Bl][l][0.4][0]{ $v_2/$ GeV}
\psfrag{BB}[][l][0.4][90]{ $m/$ GeV}
\includegraphics[scale=0.4]{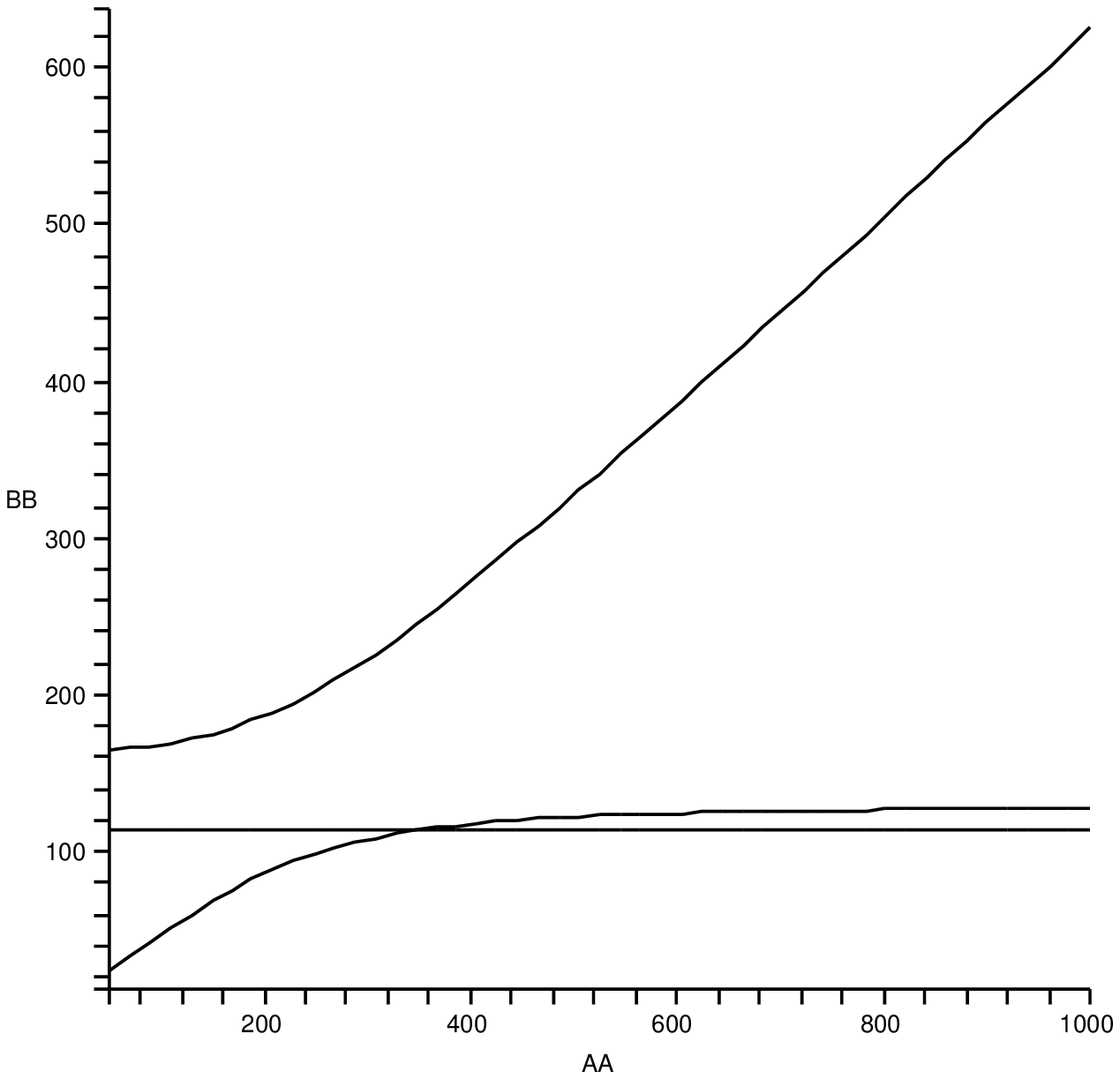}
\psfrag{AA}[Bl][l][0.4][0]{ $v_2/$ GeV}
\psfrag{BB}[][l][0.4][90]{ $m/$ GeV}
\includegraphics[scale=0.4]{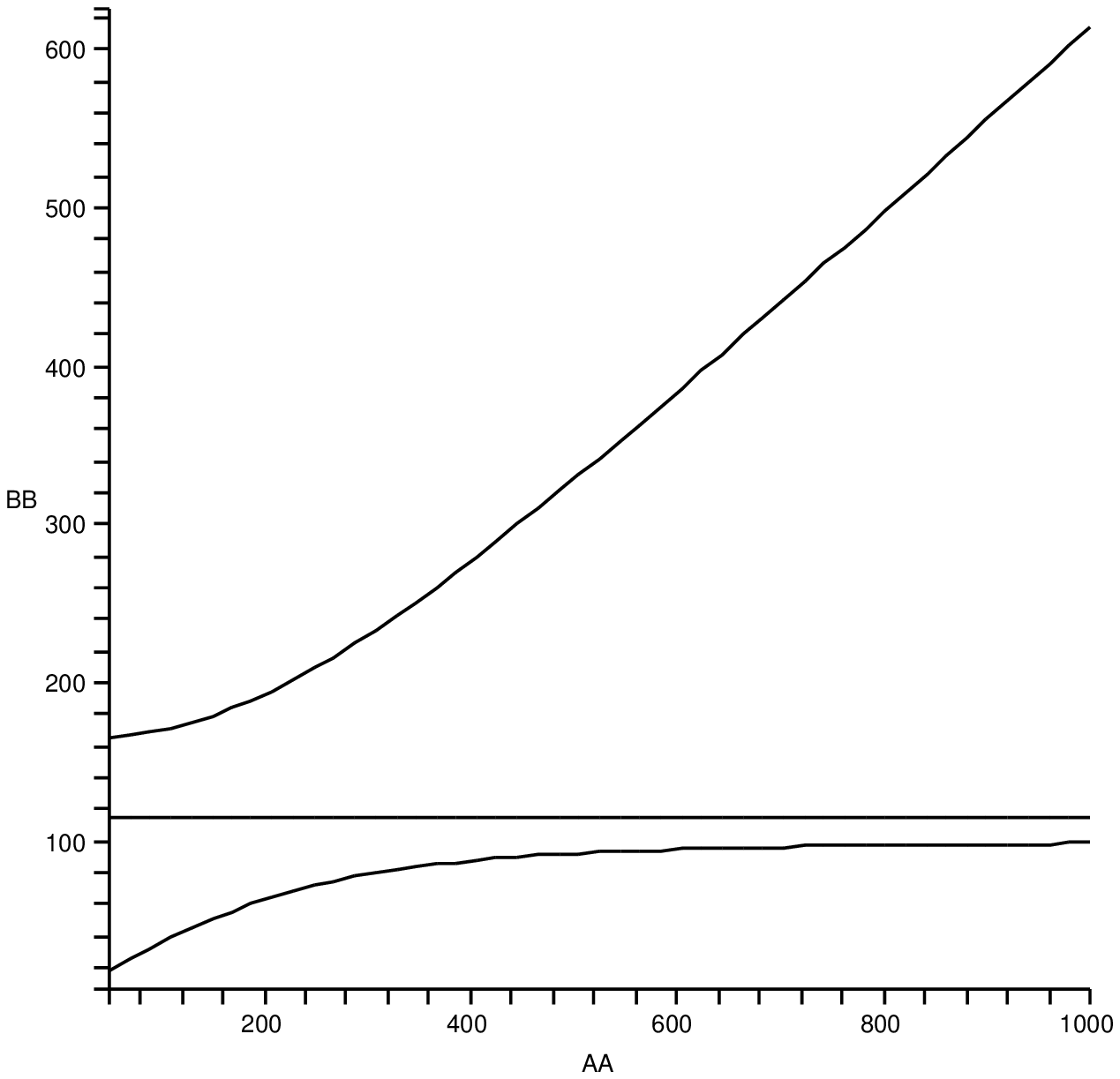}
\end{center}
\caption{Mass eigenvalues of the scalar fields in dependence of
 $v_2 = \sqrt{2} \, |\langle \varphi \rangle|$. For the $U(1)_X$ 
 gauge coupling we choose $g_4(m_Z)=0.3$ (left) and 
$g_4(m_Z)=0.7$ (right). From the $W$-mass follows  
$\sqrt{2} |\langle H \rangle|= 246$ GeV.}
\end{figure}

\section{Outlook}

We conclude this paper with a ``to-do-list'' of some open questions
that are worth to be addressed in the future: 
\begin{itemize}
\item Full classification of models beyond the SM in NCG
(LHC signature and/or dark matter?)
\item Mechanisms for Neutrino masses 
(Dirac/Majorana masses or something  different?)
\item Renormalisation group flow for all couplings in the spectral action 
(Exact Renorm Groups, M. Reuter et al.)
\item Spectral triples with Lorentzian signature 
(A. Rennie, M. Paschke, R. Verch,...)
\end{itemize}

\section*{Acknowledgments}
The author wishes  to thank Thomas Sch{\"u}cker, Jan-Hendrik Jureit and
Bruno Iochum for their collaboration. This work is funded by the 
Deutsche Forschungsgemeinschaft.

\end{document}